\documentclass{PoS}

\title{B-hadron production in top quark decay}

\ShortTitle{B-hadron production in top quark decay}

\author{\speaker{Gennaro Corcella}\vspace{0.1cm}\\
        Dipartimento di Fisica\\
        Universit\`a di Roma `La Sapienza', Italy\\
        P.le A.Moro 2, I-00185 Roma, Italy\\
        E-mail: \email{Gennaro.Corcella@roma1.infn.it}
\begin{flushright}
\vspace{-8.cm}
\rm{ROME1/1424/06}\\
\rm{February 2006}
\vspace{6.5cm}
\end{flushright}}

\author{Volker Drollinger\\
        Dipartimento di Fisica Galileo Galilei\\     
        Universit\`a di Padova and INFN, Sezione di Padova\\
        Via Marzolo 8, I-35131 Padova, Italy\\
        E-mail: \email{Volker.Drollinger@cern.ch}}

\abstract{We present the energy distribution of 
$b$-flavoured hadrons in top quark decay using the
PYTHIA and HERWIG event generators, which we tune to LEP and SLD data.
We find that fitting the string and cluster models 
is essential to reproduce the $e^+e^-$ data and to reliably
predict $B$-hadron production in top decay.
We also compare the PYTHIA and HERWIG results with
the ones yielded by resummed calculations based on the fragmentation
function formalism.}

\FullConference{International Workshop on Top Quark Physics\\
		 January 12-15, 2006\\
		 Coimbra, Portugal}

\begin{document}

Top quark phenomenology is one of the main
fields of investigation in theoretical and experimental particle physics.
The experiments at the Tevatron accelerator and, in future,
at the LHC will allow one to perform improved measurements of
the top properties, such as its mass, thanks to the production of
large amounts of $t\bar t$ pairs.

In this paper we study bottom quark fragmentation in top quark decay
($t\to bW$), which is responsible of one of the largest contribution to  
the uncertainty on 
the top mass measurement at the Tevatron \cite{cdf,d0} and the LHC \cite{lhc}.
In particular, in the analysis of Ref.~\cite{avto} the top quark mass 
is determined using  at the LHC
final states with leptons and $J/\psi$'s, where the leptons
come from the $W$ decay $W\to\ell\nu$, and the $J/\psi$'s from the
decay of a $b$-flavoured hadron $B$.  
In \cite{avto} the PYTHIA event generator \cite{pythia} was exploited,
and the error on $m_t$ was estimated to be 
$\Delta m_t\simeq 1$~GeV, with $b$-fragmentation being the largest source 
of uncertainty. In \cite{cms}, the invariant mass $m_{B\ell}$, yielded by the
HERWIG \cite{herwig} event generator, was used to fit $m_t$,
and the impact of matrix-element corrections to the simulation of top decay
\cite{corsey} was investigated. 

Bottom quark fragmentation in top decay was studied in \cite{cm,ccm},
following the method of perturbative fragmentation functions \cite{mele}.
The NLO $b$-quark energy spectrum is expressed as the 
convolution of a coefficient function, describing the emission of a 
massless parton, and a perturbative fragmentation function $D(m_b,\mu_F)$,
associated with the transition of a massless parton into a massive $b$.
$D(m_b,\mu_F)$ follows the Dokshitzer--Gribov--Lipatov--Altarelli--Parisi
(DGLAP) evolution equations \cite{ap,dgl}, which can be solved
once an initial condition at a scale $\mu_{0F}$ is given. 
The initial condition
of the perturbative fragmentation function, first computed in
\cite{mele}, was proved to be process-independent in \cite{cc}.
Solving the DGLAP evolution equations we can resum the large 
$\ln(m_t^2/m_b^2)$ which appears in the NLO
massive $b$-spectrum (collinear resummation).
Both the top-decay coefficient function, computed in \cite{cm}, 
and the initial condition
$D(m_b,\mu_{0F})$ present terms which become large
when the $b$-quark energy fraction
$x_b$ approaches 1, which corresponds to soft-gluon radiation.
Soft contributions in the initial condition (process independent) 
and in the coefficient function (process dependent) were resummed in
the NLL approximation in \cite{cc} and \cite{ccm}, respectively.
In order to predict the spectrum of $b$-flavoured hadrons, perturbative
calculations need to be supplement by non-perturbative models.
In \cite{cm,ccm}, the $B$-hadron spectrum in top decay
was presented, after fitting a few hadronization
models to SLD \cite{sld} and ALEPH \cite{aleph} data.

Following the lines of \cite{cv}, in this paper we would like to
address $b$-fragmentation in top decay, using the PYTHIA and HERWIG event
generators. As discussed in \cite{cv}, PYTHIA and HERWIG simulate
multiple radiation in top decay in the soft or collinear 
approximation, and are
provided with matrix-element corrections \cite{corsey,norrbin}
to allow hard and large-angle 
emission. The hadronization mechanism is simulated by the string
model \cite{string} in PYTHIA, and by the cluster model \cite{cluster} in 
HERWIG. 

For the sake of a reliable prediction of the $B$-energy distribution
in $t\to bW$, we need to use models and parametrizations which are able
to describe well the $B$-spectrum at $e^+e^-$ machines. 
We consider ALEPH \cite{aleph}, OPAL \cite{opal} and
SLD \cite{sld} data on the $B$ energy fraction $x_B$ in $Z\to b\bar b$
events, where $x_B$ is defined as follows:
\begin{equation}
x_B={{2p_B\cdot p_Z}\over{m_Z^2}},
\end{equation}
with $p_Z$ and $p_B$ being the $Z$ and $B$ momenta, respectively.

\begin{figure}[ht!]
\centerline{\resizebox{0.68\textwidth}{!}{\includegraphics{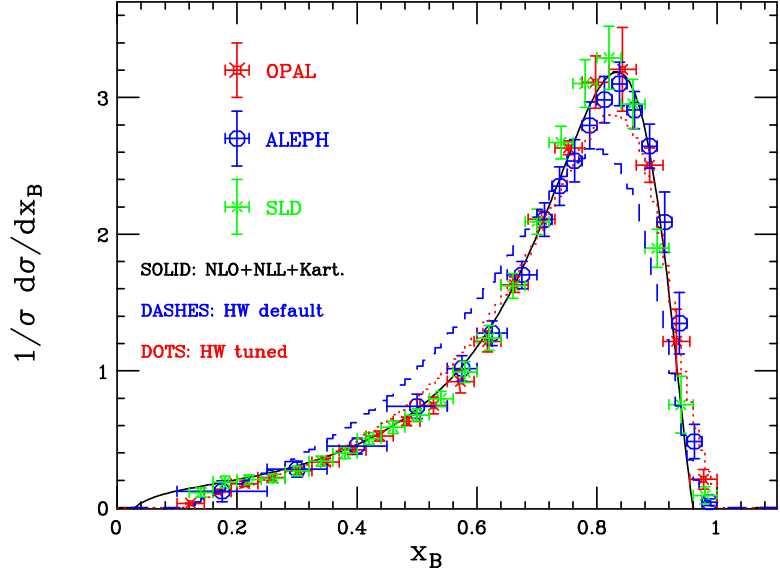}}}
\caption{Data from LEP and SLD experiments, compared with the NLO+NLL
calculation convoluted with the Kartvelishvili model (solid) 
and HERWIG 6.506, using the default parametrization (dashed)
and our tuning (dotted).}
\label{eeh}
\end{figure}

\begin{figure}[ht!]
\centerline{\resizebox{0.68\textwidth}{!}{\includegraphics{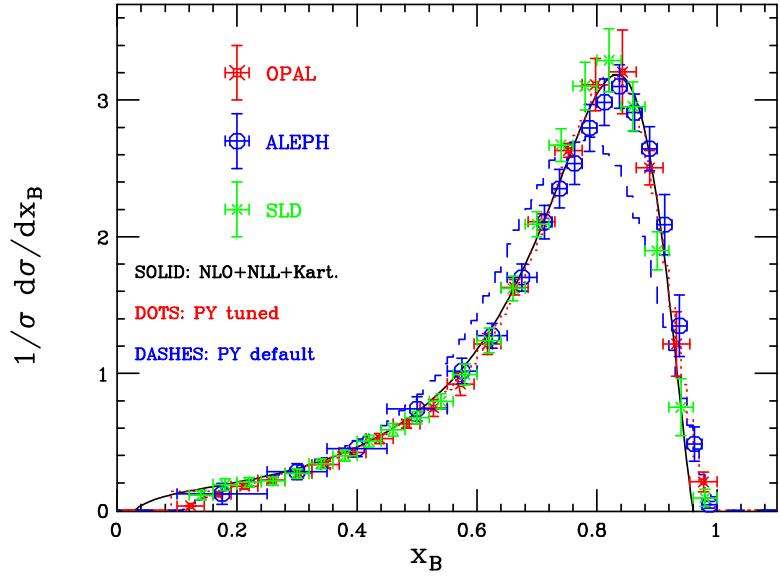}}}
\caption{As in Fig.~\protect\ref{eeh}, but comparing data and the NLO+NLL
calculation with default (dashed) and tuned (dotted) PYTHIA 6.220.}
\label{eep}
\end{figure}

We use the versions HERWIG 6.506 and PYTHIA 6.220, and find that 
the default parametrizations are unable to fit such data, as one gets 
$\chi^2/\mathrm{dof}=739.4/61$ (HERWIG) and $\chi^2/\mathrm{dof}=467.9/61$
(PYTHIA).
As in \cite{cv}, we tune the cluster and string models to the data, while
we leave unchanged the parameters of HERWIG and PYTHIA which are related to
the perturbative phase of the parton showers. Our best fits are summarized in
Table~\ref{tab1}: for PYTHIA we are able to find a parametrization which
is able to reproduce well the data ($\chi^2/\mathrm{dof}=45.7/61$);
HERWIG, even after the fit, is still
not able to describe the $x_B$-spectrum very well, but the comparison
is anyway much better than with the default parameters 
($\chi^2/\mathrm{dof}=222.4/61$). We have also checked that 
the parametrizations in Table~\ref{tab1} work well for the new model
implemented in PYTHIA 6.3 \cite{p63}.
\begin{table}[t]
\caption{\label{tab1} Parameters of HERWIG and PYTHIA
hadronization models tuned to
$e^+e^-$ data, along with the $\chi^2$ per degree of freedom.}
\begin{center}
\begin{tabular}{|c|c|}\hline
HERWIG & PYTHIA \\
\hline\hline
CLSMR(1) = 0.4  &                 \\
 \hline
CLSMR(2) = 0.3  & PARJ(41) = 0.85 \\
\hline
DECWT = 0.7     & PARJ(42) = 1.03 \\
\hline
CLPOW = 2.1     & PARJ(46) = 0.85 \\
\hline
PSPLT(2) = 0.33 &                \\
\hline
\hline
$\chi^2/\mathrm{dof}$ = 222.4/61 & $\chi^2/\mathrm{dof}$ = 45.7/61 \\
\hline
\end{tabular}
\end{center}
\end{table}
The HERWIG and PYTHIA spectra, before and after the fit, along with the 
experimental data, are presented in Figures~\ref{eeh} and \ref{eep}.
For the sake of comparison, we also show the $x_B$-spectrum yielded by the 
NLO+NLL calculation of Ref.~\cite{cc}, convoluted with the Kartvelishvili
hadronization model \cite{kart}: 
\begin{equation}
D^{\mathrm{np}}(x;\gamma)=(1+\gamma)(2+\gamma) (1-x) x^\gamma.
\label{kk}
\end{equation}
We fit the model (\ref{kk}) to the data in the range 
$0.18\leq x_B\leq 0.94$, to avoid the regions at small and large $x_B$,
where, as pointed out in \cite{cc}, the resummed calculation yields
a negative distribution.
Setting
$m_Z=91.118$~GeV, $m_b=$~5 GeV and $\Lambda^{(5)}_{\overline{\mathrm{MS}}}
=200$~GeV in the perturbative calculation, we get $\gamma=17.178\pm 0.303$
From Figs.~\ref{eeh} and \ref{eep} we learn that
default HERWIG and PYTHIA are far from the data. After the
tuning, PYTHIA reproduces the data quite well, while HERWIG yields a broader
distribution; the resummed calculation is consistent with the data.
and $\chi^2/\mathrm{dof}=46.2/53$ from the fit.


\begin{figure}[ht!]
\centerline{\resizebox{0.68\textwidth}{!}{\includegraphics{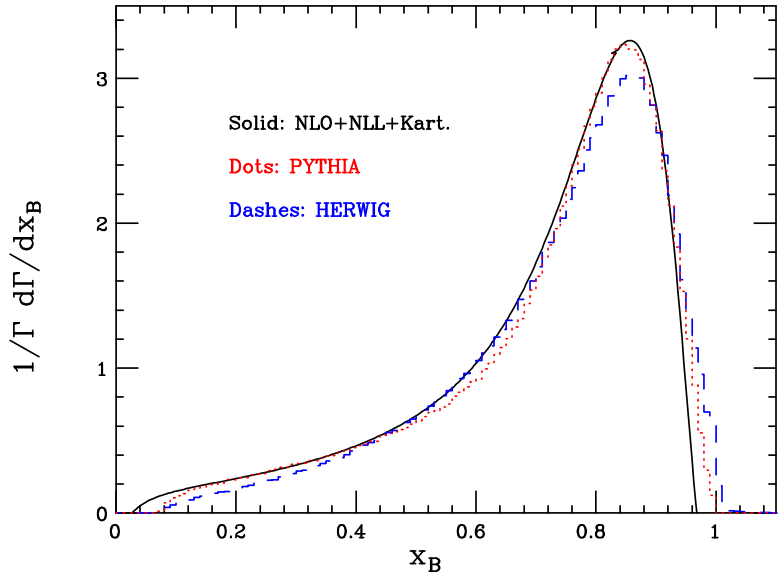}}}
\caption{$B$-hadron spectra in top decay, for $m_t=175$~GeV,
according to a NLO+NLL computation convoluted with the Kartvelishvili model
(solid line), HERWIG (dashed) and PYTHIA (dotted).}
\label{hertop}
\end{figure}

Using the parametrizations in Table~\ref{tab1}, we can predict the
$B$-energy distribution in $t\to bW$, which will be expressed in terms
of the variable
\begin{equation}
x_B={1\over {1-w}}{{2p_B\cdot p_t}\over{m_t^2}},
\end{equation}
where $p_t$ is the top momentum  and $1/(1-w)$ is a normalization factor, 
with $w=1-m_W^2/m_t^2+m_b^2/m_t^2$.
In Fig.~\ref{hertop} we present the $B$-spectrum in top decay according to
HERWIG, PYTHIA and the resummed calculation of \cite{cm,ccm}, convoluted
with the Kartvelishvili model.
The comparison exhibited in Fig.~\ref{hertop} shows similar features
to Figures~\ref{eeh} and \ref{eep}, and reflects the quality of the
fits to the $e^+e^-$ data: PYTHIA reproduces the 
peak of the resummed calculation rather well, while it is below the NLL
prediction at $x_B<0.7$, and
above at $x_B>0.9$. HERWIG is below the resummed spectrum in most the
$x_B$-range, and above it only at large $x_B$.

Finally, we wish to present results in Mellin moment space, making use
of the data reported by the DELPHI Collaboration in \cite{delphi}
on the moments of the $B$ cross section in $e^+e^-$ processes.
From the point of view of resummed calculations, 
working in moment space presents several advantages:
in $N$-space, convolutions become ordinary products and the relation between
parton- and hadron-level moments is just
$\sigma_N^B=\sigma_N^bD_N^{\mathrm{np}}$, 
where  $D_N^{\mathrm{np}}$ is the 
non-perturbative fragmentation function in Mellin space. Therefore,
there is no
need to introduce a functional form for the hadronization model
in $x_B$-space. Also, $N$-
spectra are well defined, and do not present the problems of the
$x_B$-results, which are negative at small or large $x_B$. 

In Table~\ref{tab2} we compare the DELPHI moments with the ones given
by the tuned versions of HERWIG and PYTHIA, and the
predictions for top decay. We also quote the results yielded by the NLL
calculations of Refs.~\cite{cm,ccm}, 
extracting  $D_N^{\mathrm{np}}$ from the data.
As for $e^+e^-\to b\bar b$ processes, the moments
given by HERWIG and PYTHIA 
are consistent, within the error ranges, with the DELPHI ones. 
It is interesting that HERWIG is
compatible with the DELPHI moments, even though it was only marginally 
consistent with LEP and SLD data in $x_B$-space. 
The results for top decay have similar features to the $x_B$-spectra:
PYTHIA is very close to the NLL calculation, which uses
$D_N^{\mathrm{np}}$ extracted from the DELPHI data, while
HERWIG, whose predictions are shifted toward larger $x_B$,
gives larger moments. 

In summary, we reviewed recent results on $b$-flavoured hadron
production in top quark decay. We tuned HERWIG and PYTHIA to LEP and SLD data
and presented results on the $B$-hadron spectrum in top decay in both
$x_B$ and moment spaces. In fact, fitting the cluster and string model
turned out to be necessary to reproduce the $e^+e^-$ data.
The results were also compared with resummed 
calculations, based on the fragmentation function formalism.
We think that our analysis and fits can be useful to improve the present
understanding of $b$-quark fragmentation in top quark decay.
It will be very interesting to investigate how the tuned versions of 
HERWIG and 
PYTHIA fare with respect to other observables. For example, it
may be worthwhile
reconsidering the studies in Refs.~\cite{avto,cms} with the parametrizations
which we have proposed, and estimate the contribution of $b$-fragmentation
to the uncertainty on the top quark mass reconstruction.

\begin{table}[ht!]
\caption{\label{tab2}\small  Moments
$\sigma^B_N$ from
DELPHI~\protect\cite{delphi}, and moments
in $e^+e^-$ annihilation and top ($t$) decay, 
using NLL resummed calculations, HERWIG (HW) and PYTHIA (PY).}
\small
\begin{tabular}{| c | c c c c |}
\hline
& $\langle x\rangle$ & $\langle x^2\rangle$ & $\langle x^3\rangle$
& $\langle x^4\rangle$ \\
\hline
\hline
$e^+e^-$ data $\sigma_N^B$&0.7153$\pm$0.0052 &0.5401$\pm$0.0064 &
0.4236$\pm$0.0065 &0.3406$\pm$0.0064  \\
\hline
\hline
$e^+e^-$ NLL $\sigma_N^b$   & 0.7801 & 0.6436 & 0.5479 & 0.4755  \\
\hline
$D^{\mathrm{np}}_N$ & 0.9169 & 0.8392 & 0.7731 & 0.7163 \\
\hline
$e^+e^-$ HW $\sigma_N^B$   & 0.7113 & 0.5354 & 0.4181 & 0.3353  \\
\hline
$e^+e^-$ PY $\sigma_N^B$   & 0.7162 & 0.5412 & 0.4237 & 0.3400  \\
\hline
\hline
$t$-dec. NLL $\Gamma^b_N$ & 0.7883 & 0.6615 & 0.5735 & 0.5071 \\
\hline
$t$-dec. NLL $\Gamma^B_N=\Gamma^b_N
D_N^{\mathrm{np}}$ & 0.7228 & 0.5551 & 0.4434 & 0.3632 \\
\hline
$t$-dec. HW $\Gamma^B_N$ & 0.7325 & 0.5703 & 0.4606 & 0.3814 \\
\hline
$t$-dec. PY $\Gamma^B_N$ & 0.7225 & 0.5588 & 0.4486 & 0.3688 \\
\hline
\end{tabular}
\end{table}


\begin{thebibliography}{99}
\bibitem{cdf}
CDF Collaboration, A. Abulencia et al., Phys. Rev. Lett. 96 (2006) 022004.
\bibitem{d0}
D0 Collaboration, V.M. Abazov et al., Phys. Lett. B 606 (2005) 25.
\bibitem{lhc}
M. Beneke, I. Efthymiopoulos, M.L. Mangano, J. Womersley et al., 
in Proceedings of 1999 CERN Workshop on
Standard Model Physics (and more) at the LHC, CERN 2000-004, 
G. Altarelli and M.L. Mangano eds., p.~419, hep-ph/0003033.
\bibitem{avto}
A. Kharchilava, Phys. Lett. B 476 (2000) 73.
\bibitem{pythia}
T. Sj\"ostrand, L. L\"onnblad and S. Mrenna, hep-ph/0108264.
\bibitem{cms}
G. Corcella, J.\ Phys.\ G26 (2000) 634;\\
G. Corcella, M.L. Mangano and M.H. Seymour, JHEP 0007 (2000) 004. 
\bibitem{herwig}
G. Corcella, I.G. Knowles, G. Marchesini, S. Moretti, K. Odagiri,
P. Richardson, M.H. Seymour, B.R. Webber, JHEP 0101 (2001) 010.
\bibitem{corsey}
G. Corcella and M.H. Seymour, Phys. Lett. B442 (1998) 417.
\bibitem{cm}
G. Corcella and A.D. Mitov, Nucl. Phys. B623 (2002) 247.
\bibitem{ccm}
M. Cacciari, G. Corcella and A.D. Mitov, JHEP  0212  (2002) 015.
\bibitem{mele}
B. Mele and P. Nason, Nucl. Phys. B   361 (1991) 626.
\bibitem{cc}
M. Cacciari and S. Catani, Nucl.  Phys.  B617 (2001) 253.
\bibitem{ap}
G. Altarelli and G. Parisi, Nucl. Phys. B126 (1977) 298.
\bibitem{dgl}
L.N. Lipatov, Sov. J. Nucl. Phys. 20 (1975) 95;
V.N. Gribov and L.N. Lipatov, Sov. J. Nucl. Phys. 15 (1972) 438;
Yu.L. Dokshitzer, Sov. Phys. 46 (1977) 641.
\bibitem{sld}
SLD Collaboration, K. Abe et al., Phys. Rev. Lett. 84 (2000) 4300.
\bibitem{aleph}
ALEPH Collaboration, A. Heister et al., Phys. Lett. B512 (2001) 30.
\bibitem{cv}
G. Corcella and V. Drollinger, Nucl. Phys. B730 (2005) 82.
\bibitem{norrbin}
E. Norrbin and T. Sjostrand, Nucl. Phys. B603 (2001) 297.
\bibitem{string}
B. Andersson, G. Gustafson, G. Ingelman, T. Sj\"ostrand,
Phys. Rept. 97 (1983) 31.
\bibitem{cluster}
B.R. Webber, Nucl. Phys. B238 (1984) 492.
\bibitem{opal}
OPAL Collaboration, G. Abbiendi et al., Eur. Phys. J. C29 (2003) 463.
\bibitem{p63}
T. Sj\"ostrand, L. L\"onnblad, S. Mrenna and P.Z. Skands, hep-ph/0308153.
\bibitem{kart}
V.G. Kartvelishvili, A.K. Likehoded and V.A. Petrov, Phys. Lett. B78 (1978)
615. 
\bibitem{delphi}
DELPHI Collaboration, G. Barker et al., DELPHI 2002-069, CONF 603.
\end{thebibliography}
\end{document}